\documentclass[superscriptaddress,amsfonts,amssymb,amsmath,twocolumn]{revtex4}

\usepackage{epsfig}
\usepackage{graphics}  
\usepackage{amsmath}
\usepackage{xcolor}
\date{\today}

\begin{document}


\title{The Unruh effect for mixing neutrinos}
    

\author{Gabriel Cozzella}\email{cozzella@ift.unesp.br}    
\affiliation{Instituto de F\'\i sica Te\'orica, 
Universidade Estadual Paulista, Rua Dr.\ Bento Teobaldo Ferraz, 271, 01140-070, S\~ao Paulo, S\~ao Paulo, Brazil}

\author{Stephen A. Fulling}\email{fulling@math.tamu.edu}    
\affiliation{Department of Physics, Texas A\& M University, College Station, Texas, 77843-4242, USA}
\affiliation{Department of Mathematics, Texas A\& M University, College Station, Texas, 77843-3368, USA}

\author{Andr\'e G.\ S.\ Landulfo}\email{andre.landulfo@ufabc.edu.br}
\affiliation{Centro de Ci\^encias Naturais e Humanas,
Universidade Federal do ABC, Avenida dos Estados, 5001, 09210-580, Santo Andr\'e, S\~ao Paulo, Brazil}

\author{George E.\ A.\ Matsas}\email{matsas@ift.unesp.br}
\affiliation{Instituto de F\'\i sica Te\'orica, 
Universidade Estadual Paulista, Rua Dr.\ Bento Teobaldo Ferraz, 271, 01140-070, S\~ao Paulo, S\~ao Paulo, Brazil}

\author{Daniel A.\ T.\ Vanzella}\email{vanzella@ifsc.usp.br}
\affiliation{Instituto de F\'\i sica de S\~ao Carlos,
Universidade de S\~ao Paulo, Caixa Postal 369, 13560-970, S\~ao Carlos, S\~ao Paulo, Brazil}

\pacs{} 
    

\begin{abstract}
Recently, the inverse $\beta$-decay rate calculated with respect to uniformly accelerated observers (experiencing the Unruh thermal bath) was revisited. Concerns have been raised regarding the compatibility of inertial and accelerated observers' results when neutrino mixing is taken into account. Here, we show that these concerns are unfounded by discussing the properties of the Unruh thermal bath with mixing neutrinos and explicitly calculating the decay rates according to both sets of observers,  confirming thus that they are in agreement. The Unruh effect is perfectly valid for mixing neutrinos.
\end{abstract}


\maketitle

\section{Introduction}\label{sec:I} 

The Unruh effect, which states that uniformly accelerated observers with proper acceleration $a$  perceives the Minkowski vacuum as a thermal state with temperature $T_U\equiv a/2\pi$, was initially derived assuming free quantum fields~\cite{U76}. Later, it was shown to be valid also for interacting ones~\cite{BW76,S82,GP76,GP78}. Surprisingly, perhaps, only recently the Unruh effect has been discussed in the context of mixing neutrinos, with disturbing conclusions being drawn. In Ref.~\cite{T16}, it is claimed that the inverse $\beta$-decay rate for an accelerated proton as calculated with respect to inertial and uniformly accelerated observers (experiencing the Unruh thermal bath) would disagree with respect to each other when taking into account the existence of multiple families of mixing neutrinos. We claim that this is impossible because calculations of observables must necessarily yield the same answer regardless of the frame used in intermediate steps. Thus, either the Unruh effect is wrong (contradicting several previous results~\cite{CHM08}, including what we consider to be a virtual observation of it~\cite{CLMV17}) or some mistake was made in the previously mentioned analysis. The purpose of this paper is to discuss the Unruh thermal bath for mixing neutrino fields and also revisit the inverse $\beta$ decay for accelerated protons with the aim to show that the Unruh effect is perfectly valid in this setting.

The paper is organized as follows. In Sec.~\ref{sec:II}, we discuss the Unruh thermal bath for the case of mixing neutrinos with particular attention to when we can consider flavor particle states as legitimate quantum states and how this is reflected    in measurements made involving the Unruh thermal bath. In Sec.~\ref{sec:III} we set the stage for calculating the inverse $\beta$-decay rate for accelerated protons. Sec.~\ref{sec:IV} concerns the calculation of the inverse $\beta$-decay rate from the inertial point of view. In Sec.~\ref{sec:V}, we calculate independently the inverse $\beta$-decay rate from the point of view of uniformly accelerated observers and show that the result is in full agreement with the one previously obtained in Sec.~\ref{sec:IV}. Our closing remarks appear in Sec.~\ref{sec:VI}. 

Throughout this work we use $(+,-,-,-)$ signature for the Minkowski metric, $\eta_{\mu \nu}$, and natural units, $\hbar=c=k_{\rm B}=1$, unless stated otherwise. The same conventions as in Ref.~\cite{S03} are followed for the Dirac matrices and normal modes.


\section{The Unruh effect for mixing neutrinos}\label{sec:II}

In this section, we will analyze some properties of the Unruh thermal bath assuming the existence of mixing neutrino fields $\hat{\nu}_i$, $i \in \{1,2,3\}$, each with mass $m_i$. For this purpose, let us begin by setting our notation and briefly reviewing some relevant features of the Unruh effect for fermionic fields. 

\subsection{The Unruh effect for fermionic fields}

Consider a fermionic field $\hat{\psi}$ with mass $m$ satisfying Dirac's equation. Inertial observers following the orbits of the timelike Killing field $\partial_t$, where $\{t,x,y,z\}$ are usual Cartesian coordinates covering  Minkowski spacetime, expand $\hat{\psi}$ in terms of positive- and negative-frequency modes (with respect to $\partial_t$), $u^{+\omega}_{\vec{k},\sigma}$ and $u^{-\omega}_{\vec{k},\sigma},$ respectively, as
\begin{equation}
    \hat{\psi} = \sum_{\sigma=\pm} \int d^3 {k} \left( \hat{a}_{\vec{k},\sigma} u^{+\omega}_{\vec{k},\sigma}+\hat{b}^{\dagger}_{\vec{k},\sigma} u^{-\omega}_{\vec{k},-\sigma} \right), \label{field_expansion}
\end{equation}
where $\omega\equiv\sqrt{|\vec{k}|^2+m^2}$, $\vec{k} = (k^x,k^y,k^z)\in\mathbb{R}^3$, and $\sigma\in \{+,-\}.$ The modes $u^{\pm\omega}_{\vec{k},\sigma}$ are given by
\begin{equation}
    u^{\pm \omega}_{\vec{k},\sigma} = \frac{e^{\mp i k_{\mu} x^\mu}}{(2 \pi)^{3/2}} v^{\pm \omega}_{\sigma}(\vec{k}), \label{inertial_modes}
\end{equation}
where  $k^{\mu} = (\omega,\vec{k})$, $x^\mu = (t,x,y,z)$, and
\begin{equation}
    v^{\pm \omega}_{\sigma}(\vec{k}) = \frac{ \left( k_{\mu} \gamma^\mu \pm m \mathbb{I} \right)}{\sqrt{\left[2\omega(\omega \pm m)\right]}} \hat{v}_{\sigma},
\end{equation}
with $\gamma^\mu=\left(\gamma^0,\gamma^1,\gamma^2,\gamma^3 \right)$ being the Dirac matrices and
\begin{align}
    \hat{v}_+ \equiv
        \begin{bmatrix}
            \ 1 \ \\ 
            \ 0 \ \\
            \ 0 \ \\
            \ 0 \
        \end{bmatrix}, \ \ 
    \hat{v}_- \equiv    
        \begin{bmatrix}
            \ 0 \ \\ 
            \ 1 \ \\
            \ 0 \ \\
            \ 0 \
        \end{bmatrix}.
\end{align}
The modes are orthonormalized according to the inner product
\begin{eqnarray}
    (\psi,\phi) \equiv  \int_{\Sigma} d \Sigma_\mu \ \overline{\psi} \ \gamma^\mu \phi, \label{inner_product}
\end{eqnarray}
where $d\Sigma_\mu\equiv d\Sigma \; n_\mu,$ $d\Sigma$ is the proper-volume element on the Cauchy surface $\Sigma$,  $n^\mu$ is a future-pointing unit vector field orthogonal to $\Sigma,$ and $\overline{\psi} \equiv \psi^\dagger \gamma^0$. The fermionic annihilation and anti-fermionic creation operators, $\hat{a}_{\vec{k},\sigma}$ and $\hat{b}^{\dagger}_{\vec{k},\sigma}$, respectively,  satisfy the usual anti-commutation relations:
\begin{eqnarray}
    \big{\{}\hat{a}_{\vec{k},\sigma},\hat{a}^{\dagger }_{\vec{k}',\sigma'}\big{\}} &=& \delta^3(\vec{k}-\vec{k}')\delta_{\sigma,\sigma'}, \label{acr1} \\
    \big{\{}\hat{b}_{\vec{k},\sigma},\hat{b}^{\dagger }_{\vec{k}',\sigma'}\big{\}} &=&  \delta^3(\vec{k}-\vec{k}')\delta_{\sigma,\sigma'}, \label{acr2}
\end{eqnarray}
with all the other anti-commutators vanishing. We define the Minkowski vacuum state, $| 0_{\text{M}} \rangle$, as the state annihilated by all annihilation operators, i.e., 
\begin{eqnarray}
    a_{\vec{k},\sigma} | 0_{\text{M}} \rangle = b_{\vec{k},\sigma} | 0_{\text{M}} \rangle = 0, \hspace{0.4cm} \forall \vec{k}, \sigma.
\end{eqnarray} 

On the other hand, uniformly accelerating (Rindler) observers covering  the right Rindler wedge portion of Minkowski spacetime, $z > |t|$, quantize the field using a different set of normal modes more appropriate to them. In order to describe this quantization, it is convenient to cover the right Rindler wedge with coordinates $(v,x,y,u)$ in which case the line element is written as 
\begin{equation}
ds^2 = u^2 dv^2 -dx^2 -dy^2 -du^2,
\end{equation}
where $v\in (-\infty,\infty)$ and $u\in(0,\infty)$  are given by
\begin{eqnarray}
v &=& \tanh^{-1}(t/z), \\
u &=& \sqrt{z^2-t^2}. 
\end{eqnarray}
Rindler observers, which are labeled by constant values of $u, x,$ and $ y$, expand $\hat{\psi}$ (in the right Rindler wedge) in terms of positive- and negative-frequency modes (with respect to $\partial_v$), $ g^{+\varpi}_{\vec{k}_{\bot},\sigma}$ and $g^{-\varpi}_{\vec{k}_{\bot},-\sigma},$ respectively, as

\begin{align}
    \hat{\psi} &&= \sum_{\sigma = \pm} \int_0^{\infty} d\varpi \int d^2 k_{ \bot} \left[ \hat{c}_{\varpi,\vec{k}_{\bot},\sigma}  g^{+\varpi}_{\vec{k}_{\bot},\sigma} \right. \nonumber \\
    &&\left. +\hat{d}^{\dagger}_{\varpi,\vec{k}_{\bot},\sigma} g^{-\varpi}_{\vec{k}_{\bot},-\sigma} \right],  \label{field_expansion_rindler}
\end{align}
where $\varpi\in [0,\infty)$ stands for the Rindler frequency and $\vec{k}_{\bot}\equiv (k^x,k^y)\in \mathbb{R}^2$ labels the transverse momentum quantum number. The modes $ g^{\pm \varpi}_{\vec{k}_{\bot},\sigma}$,  orthonormalized according to Eq.~(\ref{inner_product}), have the form
\begin{equation}\label{rindler_modes}
 g^{\pm \varpi}_{\vec{k}_{\bot},\sigma} = \frac{e^{\mp i \varpi v/a+i \vec{k}_{\bot} \cdot \vec{x}_\bot}}{(2 \pi)^{3/2}} h_{ \sigma}(\pm \varpi,\vec{k}_{\bot}),  
\end{equation}
where $\vec{x}_\bot=(x,y)$ and 
\begin{align}
    &&h_{\sigma}(\pm \varpi,\vec{k}_{\bot}) =
    \left[\frac{ \cosh \left(\varpi \pi/a\right)}{\pi a \, l} \right]^{1/2} 
    \nonumber \\
    &&\times \gamma^0 \left[ \left(\vec{k}_{\bot} \cdot \vec{\gamma}_\bot +  m \mathbb{I} \right) K_{\pm i \varpi/a +1/2} (l \, u) \right. \nonumber \\ 
    &&\left. + i l \gamma^3 K_{\pm i \varpi/a-1/2}(l \, u) \vphantom{\left(\vec{k}_\bot \cdot \vec{\gamma}_\bot + m \right)}  \right] \hat{h}_{\sigma}, 
\end{align} 
with
\begin{align}
    \hat{h}_+ \equiv 
        \begin{bmatrix}
            \ 1 \ \\ 
            \ 0 \ \\
            \ 1 \ \\
            \ 0 \
        \end{bmatrix}, \ \ 
    \hat{h}_- \equiv    
        \begin{bmatrix}
            \ 0 \ \\ 
            \ 1 \ \\
            \ 0 \ \\
            \ -1 \
        \end{bmatrix}, 
\end{align}
$l \equiv \sqrt{|\vec{k}_{\bot}|^2+m^2},$ $\vec{\gamma}_\bot \equiv \left(\gamma^1,\gamma^2 \right)$, and  $a$ being the proper acceleration of fiducial observers labelled by $u=1/a$, with respect to whom the quantization is performed. Rindler's fermionic annihilation, $\hat{c}_{\varpi,\vec{k}_{\bot},\sigma}$,  and  anti-fermionic creation, $\hat{d}^{\dagger}_{\varpi,\vec{k}_{\bot},\sigma}$,  operators satisfy the anti-commutation relations:
\begin{eqnarray}
  &&  \big{\{}\hat{c}_{\varpi,\vec{k}_{\bot},\sigma},\hat{c}^{\dagger }_{\varpi',\vec{k}'_{\bot},\sigma'}\big{\}} = \delta(\varpi-\varpi')\delta^2(\vec{k}_{\bot}-\vec{k}'_{\bot})\delta_{\sigma,\sigma'}, \label{acr3}  \nonumber \\ \\
   && \big{\{}\hat{d}_{\varpi,\vec{k}_{\bot},\sigma},\hat{d}^{\dagger }_{\varpi',\vec{k}'_{\bot},\sigma'}\big{\}} = \delta(\varpi-\varpi')\delta^2(\vec{k}_{\bot}-\vec{k}'_{\bot})\delta_{\sigma,\sigma'}, \label{acr4} \nonumber \\
\end{eqnarray}
with all the other anti-commutators vanishing. The Rindler vacuum is the state $| 0_{\text{R}} \rangle$ defined by
\begin{eqnarray}
    \hat{c}_{\varpi,\vec{k}_{\bot},\sigma} | 0_{\text{R}} \rangle = \hat{d}_{\varpi,\vec{k}_{\bot},\sigma} | 0_{\text{R}} \rangle = 0, \hspace{0.3cm} \forall \varpi, \vec{k}_{\bot},\sigma. \ \
\end{eqnarray}

By relating Minkowski and Rindler modes, Eqs.~(\ref{inertial_modes}) and~(\ref{rindler_modes}), respectively, in the usual manner via a Bogolubov transformation~\cite{note}, the  Minkowski vacuum state, $| 0_{\text{M}} \rangle$, as  seen by Rindler observers restricted to the right Rindler wedge can be written as
\begin{eqnarray}
    \hat{\rho}_{\beta_U} &=& \bigotimes_{\varpi,\vec{k}_{\bot},\sigma,J} Z_{\varpi} \sum_{n_J=0}^1  \exp{\left(-2\pi n_J \varpi/a \right )} \nonumber \\
    &\times& |n_J; \varpi,\vec{k}_{\bot},\sigma \rangle \langle n_J;\varpi,\vec{k}_{\bot},\sigma| \label{thermal_bath}, 
\end{eqnarray}
where $Z^{-1}_{\varpi} = 1+\exp{\left(-2\pi \varpi/a \right )}$, $J=c,d$ label particles ($c$) and anti-particles ({$d$):
$$|n_c; \varpi,\vec{k}_{\bot},\sigma \rangle \equiv {c^{\dagger}}_{\varpi,\vec{k}_{\bot},\sigma}^{n_c}|0_{\text{R}}\rangle, $$ 
and 
$$|n_d; \varpi,\vec{k}_{\bot},\sigma \rangle \equiv {d^{\dagger}}_{\varpi,\vec{k}_{\bot},\sigma}^{n_d}|0_{\text{R}}\rangle.$$ We see that $\hat{\rho}_{\beta_U}$ is a thermal state at inverse temperature $\beta_U = 2\pi/a$, clearly showing the Unruh effect. 

Now, let us take our fermionic field $\hat{\psi}$ to be one of the massive neutrino fields $\hat{\nu}_i$. (No mixing appear at this point because we are neither considering interactions nor flavor neutrinos yet.) Then, it follows directly from Eq.~(\ref{thermal_bath}) that the mean flux of neutrinos with well-defined mass $m_i$, energy $\varpi$, transverse momentum $\vec{k}_\bot$, and spin $\sigma$, as  seen by Rindler observers, is given by
\begin{eqnarray}
    \bar{n}(\varpi,\vec{k}_\bot,\sigma) &\equiv& \frac{d}{d\tau}\frac{d^2}{d^2 x_\bot} \lim_{\begin{matrix}
\varpi' \to \varpi\\ 
 \vec{k}'_\bot \to \vec{k}_\bot
\end{matrix}} \Big{\langle} c^{\dagger i}_{\varpi',\vec{k}'_{\bot},\sigma} c^{i}_{\varpi,{\vec{k}}_{i \bot},\sigma} \Big{\rangle}_{\hat{\rho}_{\beta_U}} \nonumber \\
    &=& (2\pi)^{-3}  n_F(\varpi), 
\end{eqnarray}
where 
\begin{equation}
    n_F(\varpi) \equiv (1+e^{{\beta_U} \varpi})^{-1},
\end{equation} 
and $\tau=v/a$ is the proper time of fiducial Rindler observers (at $u = 1/a$). As expected, the result is proportional to the Fermi-Dirac factor $n_F(\varpi)$ and only depends on the Rindler energy $\varpi$. In particular, there is no dependency on the neutrino mass. This does not mean, however, that detectors carried by a given Rindler observer and sensitive to neutrinos $\nu_i$ with different masses $m_i$,  would respond in the same way to the Unruh thermal bath.

\subsection{Fermionic particle detector}

To illustrate this point, let us define a generalization of the Unruh-DeWitt detector which couples to the neutrino field $\hat{\nu}_i$ through the interaction action
\begin{eqnarray}
    \hat{S}_D &\equiv& \lambda \sum_{i} \int d\tau  \left[ \hat{\overline{m}}(\tau) \hat{\nu}_i[x^\mu_D(\tau)] + \text{H. c.} \right], \label{detector}
\end{eqnarray}
where $\lambda$ is a (dimensional) constant, $x_D^\mu(\tau)$ is the detector's trajectory, and 
\begin{eqnarray}
    \hat{m}(\tau) &=& \hat{m}_0(\tau) \binom{\hat{\eta}}{\hat{\xi}},
\end{eqnarray}
with $\hat{m}_0(\tau)$ being a monopole operator such that
\begin{eqnarray}
    \langle e | \hat{m}_0(\tau) | g \rangle = e^{i \Delta E  \tau} \langle e |\hat{m}_0(0) | g \rangle.
\end{eqnarray}
Here, $\hat{\eta}$ and $\hat{\xi}$ are arbitrary bi-spinors satisfying the normalization conditions $\hat{\eta}^\dagger \hat{\eta}= \hat{\xi}^\dagger \hat{\xi} = 1$ and $\Delta E$ is the energy gap between the detector's excited, $|e\rangle$,  and unexcited, $|g\rangle$, states. (In Appendix \ref{sec:AP_A}, we analize  the behavior of this detector in  the simpler setting of an inertial thermal bath and highlight its nice features.)

The worldline of a uniformly accelerated detector with proper acceleration $a$ in  $(v,x,y,u)$ coordinates is given by
\begin{equation}\label{detwl}
    x^\mu_D(\tau) = (a \tau,0,0,1/a).
\end{equation}
For such a detector, the excitation rate, i.e., the excitation probability (with absorption of a Rindler neutrino with mass $m_i$) per detector proper time,  when the field is in the Minkowski vacuum is 
\begin{eqnarray}\label{proprate}
    \frac{d P_{\text{exc},i}}{d \tau} &=& \frac{d}{d\tau}  \sum_{\sigma = \pm} \int_0^\infty d\varpi \int d^2 k_\bot |\mathcal{A}_{\text{exc}} |^2 n_F(\varpi), \ \ \ \ \
\end{eqnarray}    
where
\begin{equation}\label{det_amp}
    \mathcal{A}_{\text{exc}} = i\langle e|\otimes \langle 0_\text{R} | \hat{S}_D|\nu_i; {\varpi,\vec{k}_\bot,\sigma}\rangle \otimes | g\rangle.
\end{equation}
Using Eqs. (\ref{detector}),~(\ref{detwl}), and~(\ref{det_amp}) in Eq.~(\ref{proprate}) yields
\begin{eqnarray}
     \frac{d P_{\text{exc},i}}{d \tau} &=&  \Lambda^2 \int_0^\infty d\varpi \ \delta(\varpi-\Delta E) e^{-\pi \varpi /a} \nonumber  \\ 
     &\times& \int d^2 k_\bot \left( l_i / a \right)  |K_{i\varpi /a+1/2}\left(l_i/a\right)|^2,
   \label{exc_rate}
\end{eqnarray}
where the constant $\Lambda^2 = |\lambda|^2 |\langle e |\hat{m}_0(0) | g \rangle|^2 / 2 \pi^3$ depends on the detector's specifics and we recall that $l_i =\sqrt{ |\vec{k}_\bot|^2+m^2_i}$. Note that the detector is only sensitive to particles with Rindler energy $\varpi = \Delta E$. 


The excitation rate  should be proportional to the local neutrino density. We can confirm this by extending the scalar-field  construction of the finite-volume particle number operator  given in  Ref.~\cite{KSAD17} (borrowed from quantum optics) to spinor fields. By making use of the inner product~(\ref{inner_product}), we define the neutrino density operator in an infinitesimal volume around the detector as

\begin{eqnarray}
    \hat{n}_{i,D} &\equiv& \left.\begin{matrix} d (\nu_i^+,\nu_i^+) / du \ d^2 x_\bot \end{matrix} \right|_{x^\mu(\tau)=x^\mu_D(\tau)} \label{nudensity} \\
    &=& \left.\begin{matrix} \overline{\nu}_i^+ \gamma^0 \nu_i^+ \end{matrix}\right|_{x^\mu(\tau)=x^\mu_D(\tau)},
\end{eqnarray}
where $\nu_i^+$ denotes the (Rindler) positive-frequency part of Eq.~(\ref{field_expansion_rindler}) and Eq.~(\ref{nudensity}) was evaluated over a $v=\text{const}$ surface. The expectation value of this density operator in the Minkowski vacuum $| 0_{\text{M}} \rangle$ is given by
\begin{eqnarray}
  &&  \langle 0_{\text{M}} | \hat{n}_{i,D} | 0_{\text{M}} \rangle = \frac{1}{2 \pi^4}  \int_0^\infty d \varpi \ e^{-\pi \varpi/a} \nonumber \\ 
    &\times&  \int d^2 k_\bot  \left( l_i/a \right) |K_{i\varpi/a+1/2}\left(l_i/a\right)|^2. 
    \label{nudensity_result}
\end{eqnarray}
Clearly, Eq.~(\ref{exc_rate}) is proportional to Eq.~(\ref{nudensity_result}) when restricted to particles with $\varpi=\Delta E$, confirming that the excitation rate is proportional to the local neutrino density. We also note from Eq.~(\ref{exc_rate}) that Rindler observers will have a harder time detecting more massive neutrinos since they concentrate closer to the horizon. This is in accordance with previous results obtained for the scalar field case~\cite{KSAD17,CCMV02}.

\subsection{On flavor neutrinos and the Unruh thermal bath}

Let us consider now the properties of the Unruh thermal bath in terms of flavor neutrinos. It is possible to define {\it phenomenologically} flavor states $|\nu_\alpha;\vec{k},\sigma \rangle$ and $ |\nu_\alpha;\varpi,\vec{k}_{\bot},\sigma \rangle $ for inertial and Rindler observers, respectively, in the realm of quantum field theory, where $\alpha \in \{e,\mu,\tau\}$ labels the leptonic flavor. Their usual form (see Eqs.~(\ref{flavor_states1}) and~(\ref{flavor_states2}) below) arise as a useful approximation, for instance, in the regime where the relevant momenta involved are much larger than the neutrinos masses. For more details,  we address the reader to Refs. \cite{CG, GL, BLK, CG18} (see also the final paragraph of Sec. \ref{sec:III}). From the perspective of inertial observers, this regime is given by
\begin{equation}\label{momentum_condition_in}
|\vec{k}|^2 \gg m^2_i, \ \forall i,
\end{equation}
as can be seen from Eq.~(\ref{inertial_modes}). As for the Rindler observers, Eq.~(\ref{rindler_modes}) shows us that this will be the case whenever we deal with neutrinos that have transverse momentum satisfying 
\begin{eqnarray}
    |\vec{k}_\bot|^2 \gg m_i^2, \ \forall i.
    \label{momentum_condition}
\end{eqnarray}
Then, in the regime given by Eqs.~(\ref{momentum_condition_in}) and~(\ref{momentum_condition}), we define the flavor neutrino states as
\begin{eqnarray}
    |\nu_\alpha;\vec{k},\sigma \rangle &\equiv& \sum_i U^*_{\alpha,i} |\nu_i;\vec{k},\sigma \rangle, \label{flavor_states1} \\
   |\nu_\alpha;\varpi,\vec{k}_{\bot},\sigma \rangle &\equiv& \sum_i U^*_{\alpha,i} |\nu_i;\varpi,\vec{k}_{\bot },\sigma \rangle, \label{flavor_states2}
\end{eqnarray}
where $U_{\alpha,i}$ is the PMNS matrix \cite{PMNS68}. 

In this high-momentum regime, we can take an uniformly accelerating detector and ask, for instance, what its excitation rate is for detecting an $\alpha$-flavor neutrino from the Unruh thermal bath. From Eq.~(\ref{flavor_states2}), we can see that the probability of a $\nu_i$ neutrino reaching the detector to collapse as a $\nu_\alpha$ neutrino is $| U_{\alpha,i} |^2$. As a result we find that
\begin{eqnarray}\label{flavor_prob}
     \frac{d P_{\text{exc},\alpha}}{d \tau} \left.\begin{matrix} \vphantom{ \frac{d P_{\text{exc},i}}{d \tau} }\end{matrix} \right|_{  |\vec{k}_\bot|^2 \gg m_i^2} &=&  \sum_i | U_{\alpha,i} |^2   \  \frac{d P_{\text{exc},i}}{d \tau}  \left.\begin{matrix} \vphantom{ \frac{d P_{\text{exc},i}}{d \tau} }\end{matrix} \right|_{  |\vec{k}_\bot|^2 \gg m_i^2} \nonumber \\
      &\approx& \frac{d P_{\text{exc},i}}{d \tau} \left. \begin{matrix} \vphantom{\frac{d P_{\text{exc},i}}{d \tau}} \end{matrix} \right|_{m_i=0},
\end{eqnarray}
where we have used the unitarity of the PMNS matrix in the last step.  We can see from Eq.~(\ref{flavor_prob}) that the predicted behavior of flavor neutrinos (when well-defined) is in no way deviant from what it would be expected from the thermality predicted by the Unruh effect. 

We proceed now to evaluate the inverse $\beta$-decay rate for accelerated protons with neutrino mixing from the point of view of both inertial and accelerated observers and show that they agree. This illustrates how the Unruh effect is perfectly consistent with neutrino mixing.

%

\section{Semi-classical inverse $\beta$ decay with neutrino mixing}\label{sec:III} 

For the sake of our purposes, we adopt the same approach as \cite{T16,VM01}, where the proton, 
$|p \rangle$, and neutron,  $|n \rangle$, are seen as unexcited and excited states of a two-level system with the corresponding (proper) Hamiltonian $\hat{H}$ satisfying
\begin{align}
    \hat{H}|p \rangle &= m_p | p \rangle, \\
    \hat{H}|n \rangle &= m_n | n \rangle,
\end{align}
where $m_{p(n)}$ is the proton (neutron) mass. The proton-neutron system is assumed to have a well-prescribed spacetime trajectory described by the semi-classical current:
\begin{equation}
    \hat{j}^\mu = \frac{\hat{q}(\tau)}{\sqrt{-g} u^0} u^\mu(\tau) \delta^3(\vec{x}-\vec{x}_0(\tau)), 
    \label{current}
\end{equation}
where $g=\textup{det}(\eta_{\mu \nu})$, $u^\mu(\tau)$ is the four velocity of the linearly accelerated proton-neutron system with proper time $\tau$, proper acceleration $a = {\rm const}$, and $\vec{x}_0(\tau)$ is its spatial trajectory. The monopole operator $\hat{q}(\tau)$ is defined via the Hamiltonian by
\begin{equation}
    \hat{q}(\tau) = e^{i \hat{H} \tau} \hat{q}(0) e^{-i \hat{H} \tau},
\end{equation}
where the Fermi constant, $G_F$, will be given by $G_F \equiv |\langle n | \hat{q}(0)| p \rangle|$. The leptonic fields, in turn, will be treated as quantum fields. 

The effective weak interaction action considered here is
\begin{eqnarray}
    &\hat{S}_I& = \int d^4 x \sqrt{-g}  \left(\sum_{\alpha} \hat{\bar{\nu}}_\alpha \gamma^\mu \hat{P}_L \hat{l}_\alpha \hat{j}_\mu + \text{H. c.} \right) \nonumber \\
    &=&\int d^4 x \sqrt{-g}  \left(\sum_{\alpha,i} U^*_{\alpha,i} \hat{\bar{\nu}}_i \gamma^\mu \hat{P}_L \hat{l}_\alpha \hat{j}_\mu + \text{H. c.} \right), 
    \label{interaction_action}
\end{eqnarray}
where 
\begin{equation}
    \hat{P}_L\equiv\frac{\left(\mathbb{I} - \gamma^5 \right)}{\sqrt{2}},
\end{equation}
$\gamma^5 \equiv i \gamma^0 \gamma^1 \gamma^2 \gamma^3$, and we recall that $\hat{j}^\mu$ is given by Eq.~(\ref{current}), $\alpha \in \{e,\mu,\tau\}$ labels the leptonic flavor,
\begin{equation}
    \hat{\nu}_\alpha \equiv \sum_i U_{\alpha,i} \hat{\nu}_i, \label{nu_a}
\end{equation}
and $\hat{l}_\alpha$ stand for all electrically charged leptonic fields, $\{ e^- ,\mu^-,\tau^- \}$. 

We should view the neutrino fields $\hat{\nu}_i$ with well defined mass as the fundamental ones with the PMNS matrix elements contributing to the interaction coupling constants between the mass neutrinos and other fields. Here, $\hat{\nu}_\alpha$ stands only as a shorthand notation for the particular combination of massive neutrino fields given by Eq.~(\ref{nu_a}). As mentioned previously in Sec. \ref{sec:II}, flavor states can only be defined phenomenologically. Attempts to canonically quantize the $\hat{\nu}_\alpha$ fields in terms of positive (and negative) frequency modes give annihilation (and creation) operators whose physical meaning is unclear~\cite{note2}, precluding us from defining flavor particle states as resulting from the action of creation operators on a vacuum and constructing the associated Fock space~\cite{note3}. 
For this reason, we focus only on states associated with the $\hat{\nu}_i$ fields from now on.


%

\section{Inertial calculation}\label{sec:IV} 

The inverse $\beta$-decay process, as seen by Minkowski observers, can be generically cast in the form 
\begin{equation}
    p \to n \ \bar{l}_\alpha \ \nu_i, 
    \label{beta_decay}
\end{equation}
where $l_\alpha = \{ e^- ,\mu^-,\tau^- \} $ and ${\nu}_i = \{ \nu_1, \nu_2, \nu_3 \}$. The transition amplitude associated with Eq.~(\ref{beta_decay}) is
\begin{equation}
    \mathcal{A}^{I}_{\alpha,i} = i \langle n | \otimes  \langle \bar{l}_\alpha \nu_i | \hat{S}_I | 0_{\text{M}} \rangle \otimes | p \rangle,
    \label{amplitude}
\end{equation}
where the charged leptons $l_\alpha$ and neutrinos $\nu_i$ have quantum numbers $\sigma_{\alpha\, (i)} \in \{+,-\} $ and $\vec{k}_{\alpha\, (i)}= (k^x_{\alpha\, (i)},k^y_{\alpha\, (i)}, k^z_{\alpha\, (i)})$, and $\hat S_I$ is given in Eq.~(\ref{interaction_action}). 

In usual inertial coordinates, $x^\mu=(t,x,y,z)$, current~(\ref{current}) is written as
\begin{equation}
    \hat{j}^\mu = \frac{\hat{q}(\tau)}{a z} u^{\mu}(\tau) \delta(x) \delta(y) \delta \left(z-\sqrt{t^2-a^{-2}} \right), 
    \label{inertial_current}
\end{equation}
where $u^\mu = (az(\tau),0,0,at(\tau))$ with $t(\tau)=a^{-1} \sinh{\left(a \tau \right)}$ and $z(\tau)=a^{-1} \cosh{\left(a \tau \right)}$.

The differential decay probability per momentum-space volume $dV_k = d^3 k_\alpha \ d^3 k_i$ is given by
\begin{equation}
    \frac{d P^{\ p \to \ n \ \bar{l}_\alpha \ \nu_i }}{d V_k} =  \sum_{\sigma_\alpha, \sigma_i} | \mathcal{A}^{I}_{\alpha, i} |^2,
    \label{differential_probability}
\end{equation}
allowing us to define the decay rate per momentum-space volume as
\begin{equation}
    \frac{d \Gamma^{\ p \to \ n \ \bar{l}_\alpha \ \nu_i }}{d V_k} =  \frac{1}{\Delta \tau} \frac{d P^{\ p \to \ n \ \bar{l}_\alpha \ \nu_i }}{d V_k}, 
    \label{differential_decay_rate}
\end{equation}
where $\Delta \tau$ is the total proper time along the trajectory of the proton-neutron system. 
Inserting Eqs.~(\ref{interaction_action}), (\ref{inertial_current}), and~(\ref{field_expansion}) into Eq.~(\ref{amplitude}), we write Eq.~(\ref{differential_decay_rate}) as
\begin{eqnarray}
    &&\frac{d \Gamma^{\ p \to \ n \ \bar{l}_\alpha \ \nu_i }}{dV_k} = \frac{2 G_F^2 |U_{\alpha,i}|^2}{(2\pi)^6}  \nonumber \\
    &\times&  \int_{-\infty}^{\infty} d\xi  \exp{\left( 2i [\Delta m\, \xi + a^{-1} (\omega_\alpha\omega_i) \sinh{(a \xi)} ] \right)}  \nonumber \\
    &\times& \frac{1}{\omega_\alpha \omega_i} [k^z_i k^z_\alpha  + \omega_i \omega_\alpha + F(k^x_{i,\alpha},k^y_{i,\alpha})],      
\end{eqnarray}
where $\Delta m \equiv m_n-m_p$, we have made an inverse boost in the $z$-direction [to factor out the proper time integral implicitly contained in Eq.~(\ref{differential_probability})], and $F(k^x_{i,\alpha},k^y_{i,\alpha})$ is an odd function of its arguments whose form is not important here since it will not contribute to the decay rate when integrated over $dV_k$. 

By integrating over momenta, we obtain the total decay rate
\begin{eqnarray}
    \Gamma^{\ p \to \ n \ \bar{l}_\alpha \ \nu_i } &=& \frac{G_F^2 |U_{\alpha,i}|^2}{ \pi^4 a} e^{\frac{-\pi \Delta m}{a}}  \int_{0}^{\infty} dk_\alpha \ k^2_\alpha \int_{0}^{\infty} dk_i \ k^2_i \nonumber \\
    &\times& K_{2i \Delta m /a} \left({2} \left( \omega_\alpha+\omega_i \right)/a \right),
\end{eqnarray}
where $k_{\alpha (i)} \equiv |\vec{k}_{\alpha (i)}|$. Now, by using the same complex integration procedure employed in Ref.~\cite{S03} we can rewrite the expression above as a double integral over the complex plane, i.e., 
\begin{eqnarray}
    &&\Gamma^{\ p \to \ n \ \bar{l}_\alpha \ \nu_i } = \frac{G_F^2 a^5 |U_{\alpha,i}|^2}{32\, \pi^{7/2} } e^{-{\pi \Delta m}/{a}} \nonumber \\
    &\times&  \int_{C_t} \frac{dt}{2 \pi i} \int_{C_s} \frac{ds}{2\pi i} \left| \Gamma \left(3 -s-t+{i\Delta m}/{a} \right) \right|^2 \nonumber \\
    &\times& \frac{ \Gamma(-s)\Gamma(-t) \Gamma(2-t)  \Gamma(2-s)  }{ \Gamma(3-s-t) \Gamma(7/2-s-t) }
    \left[\frac{m_\alpha}{a} \right]^{2t}    \left[\frac{m_i}{a} \right]^{2s} \!\!\!\!\! ,
    \label{decay_rate_final_form}
\end{eqnarray}
where $C_s$ and $C_t$ are integration contours containing all poles of the $\Gamma$ functions both in the $s$ and $t$ planes.

Although apparently unwieldy, this expression for the decay rate is convenient for our purposes of analytically confirming the equality between Eq.~(\ref{decay_rate_final_form}) and the analogous result obtained with respect to Rindler observers, to which we proceed now.

%

\section{Rindler calculation}\label{sec:V} 

Uniformly accelerated observers see the single inverse $\beta$-decay process considered by inertial observers, Eq.~(\ref{beta_decay}), as a set of three processes, namely
\begin{enumerate}
    \item $\ p +l_\alpha \to \ n + \nu_i$,
    \item \ $p + \bar{\nu}_i \to n + \bar{l}_\alpha$,
    \item \ $p + l_\alpha + \bar{\nu}_i \to n$,
\end{enumerate}
i.e, protons lying at rest with the Rindler observers would decay into neutrons by the absorption (and possible emission) of leptons from (to) the Unruh thermal bath. 

In the $(v,x,y,u)$ coordinate system, current~(\ref{current}) is expressed as
\begin{equation}
    \hat{j}^\mu = q u^\mu \delta(x)\delta(y)\delta(u-a^{-1})
\end{equation}
with $u^\mu = (a,0,0,0)$.

To obtain the total decay rate, we must sum incoherently processes (1)-(3). Let us outline now the procedure to calculate the decay rate specifically for process (1), since processes (2) and (3) will be similar. Firstly, we calculate the interaction amplitude by using Eqs.~(\ref{field_expansion_rindler}) and~(\ref{interaction_action}),
\begin{eqnarray}
    \mathcal{A}^{R,(1)}_{\alpha,i} &=& i \langle n | \otimes  \langle  \nu_i | \hat{S}_I | l_\alpha \rangle \otimes | p \rangle \nonumber \\
    &=& \frac{i  G_F}{(2\pi)^2 \sqrt{2}} U^*_{\alpha,i} \delta(\varpi_\alpha- \varpi_i-\Delta m ) \nonumber \\
    &\times& \left[ \bar{g}^{+\varpi_i}_{\vec{k}_{i \bot},\sigma_i} \gamma^0(\mathbb{I}- \gamma^5) g^{+\varpi_\alpha}_{\vec{k}_{\alpha \bot},\sigma_\alpha} \right],
    \label{accelerated_amplitude}
\end{eqnarray}
where the neutrino $\nu_i$ has quantum numbers $(\sigma_i,\varpi_i,\vec{k}_{i \bot})$ and the charged lepton has quantum numbers $(\sigma_\alpha,\varpi_{\alpha},\vec{k}_{\alpha \bot})$ and we recall that $\bar{g} = g^\dagger \gamma^0$. We square it to obtain the differential probability of decay per Rindler momentum-space volume $dV_{k,R} = d\varpi_{\alpha} d^2 k_{\alpha \bot} d\varpi_{i} d^2 k_{i \bot}$. 

The interaction rate, 
\begin{equation}
    \Gamma^{(1,R)} = \sum_{\sigma_\alpha,\sigma_i} \int dV_{k,R} \frac{\left|\mathcal{A}^{R,(1)}_{\alpha,i} \right|^2 }{\Delta \tau} n_F (\varpi_{\alpha}) \left[1-n_F (\varpi_{i}) \right],
\end{equation}
is obtained by dividing the differential probability by the total proper time $\Delta \tau$, multiplying it by the relevant fermionic thermal factors, and integrating over $dV_{k,R}$. 

Following a similar recipe for processes (2) and (3) we can write the total interaction rate, according to Rindler observers, as 
\begin{eqnarray}
    &&\Gamma^{\ p \to \ n \ \bar{l}_\alpha \ \nu_i ,R} = \sum_{j=1}^3 \Gamma^{(j,R)} \nonumber \\
    &=&\frac{ G_F^2 |U_{\alpha,i}|^2}{8 a^2  \pi^7 }  e^{-\pi \Delta m/a}  \int_{-\infty}^\infty d\varpi_{i} \iint dk^x_i dk^x_\alpha \nonumber \\ 
    &\times& \iint dk^y_i dk^y_\alpha \ l_i l_\alpha  
    \left |K_{{1}/{2}+i   \varpi_{\alpha}/ a}(l_\alpha/a) \right|^2\nonumber \\ 
    &\times&  \left |K_{{1}/{2}+i ({\varpi_{\alpha}-\Delta m})/{a}} (l_i/a) \right|^2.
\end{eqnarray}
    
Now, following the reasoning of Ref.~\cite{S03} we use Eqs.~(6.412) and the definition of the Meijer G-function [Eq.~(9.301)] of Ref.~\cite{GR}, along with Eq.~(5.6.66) of Ref.~\cite{BM} to write the total interaction rate according to Rindler observers as
\begin{eqnarray}
    && \Gamma^{\ p \to \ n \ \bar{l}_\alpha \ \nu_i ,R} = \frac{G_F^2 a^5     |U_{\alpha,i}|^2}{32\, \pi^{7/2} } e^{-{\pi \Delta m}/{a}} \nonumber \\
    && \times  \int_{C_t} \frac{dt}{2 \pi i} \int_{C_s} \frac{ds}{2\pi i} \left| \Gamma \left(3 -s-t+{i\Delta m}/{a} \right) \right|^2 \nonumber \\
    && \times \frac{ \Gamma(-s)\Gamma(-t) \Gamma(2-t)  \Gamma(2-s)  }{ \Gamma(3-s-t) \Gamma(7/2-s-t) } \left[\frac{m_\alpha}{a} \right]^{2t}    \left[\frac{m_i}{a} \right]^{2s}, \ \  \ \
    \label{accelerated_decay_rate_final_form}
\end{eqnarray}
which can be seen to be exactly equal to Eq.~(\ref{decay_rate_final_form}), proving our assertion that two rates coincide. 

%

\section{Conclusions}\label{sec:VI} 

We have discussed the properties of the Unruh thermal bath for mixing neutrinos and in which conditions we can legitimately speak about flavor states. Also, we have shown through an explicit calculation the equality of the decay rates for the inverse $\beta$ decay of protons as calculated by Minkowski and Rindler observers. This is not surprising, being the expected result from the general covariance of quantum field theory, but it explicitly demonstrates the importance of using the appropriate mode expansion (i.e., those that are eigenfunctions of an appropriate time-like isometry) for the neutrino fields, where all calculations are well-defined and it is completely meaningful to talk of particles. Finally, as previously stated, this also shows that there is no incompatibility between the Unruh effect and neutrino mixing, contrary to previous claims.

%

\acknowledgments

We are indebted to Vicente Pleitez for insightful comments. G.~C.\ and  A.~L., G.~M., D.~V.~were fully and partially supported by S\~ao Paulo Research Foundation (FAPESP) under Grants 2016/08025-0 and 2017/15084-6, 2015/22482-2, 2013/12165-4, respectively. G.~M.\ was also partially supported by Conselho Nacional de Desenvolvimento Cient\'\i fico e Tecnol\'ogico (CNPq).

\appendix 
\section{Detectors' behavior}\label{sec:AP_A}

In order to better understand the behavior of the fermionic detector defined in Sec.~\ref{sec:II}, we apply it to an usual inertial thermal bath (at inverse temperature $\beta$) of massive neutrinos, described by the density matrix
\begin{eqnarray}
        \hat{\rho}_{\beta} &=& \bigotimes_{\vec{k},\sigma_i,i,J} \frac{1}{1+\exp{\left(-\beta \omega_i \right )}}  \nonumber \\
    &\times& \sum_{n_{i,J}=0}^1 \exp{\left(- n_{i,J} \beta \omega_i \right )} |n_{i,J}; \vec{k},\sigma \rangle \langle n_{i,J};\vec{k},\sigma|, \nonumber \\ \label{inertial_thermal_bath}
\end{eqnarray}
where $\omega_i$ satisfies the usual dispersion relation $\omega_i=\sqrt{|\vec{k}|^2+m_i^2}$ and $J=a,b$ label particles ($a$) and anti-particles ($b$):
$$|n_{i,a}; \vec{k},\sigma \rangle \equiv {a^{\dagger}}_{ \vec{k},\sigma}^{n_{i,a}}|0_{\text{M}}\rangle, $$ 
and 
$$|n_{i,b};  \vec{k},\sigma \rangle \equiv {b^{\dagger}}_{ \vec{k},\sigma}^{n_{i,b}}|0_{\text{M}}\rangle,$$
with ${a^{\dagger}}_{ \vec{k},\sigma}$ and ${b^{\dagger}}_{ \vec{k},\sigma}$ being the fermionic and 
 anti-fermionic creation operators, respectively. By using Eq.~(\ref{detector}), we compute the mean excitation rate for an inertial detector with worldline $x_D^{\mu}(\tau)=(\tau,0,0,0)$ (in $\{t,x,y,z\}$ coordinates) due to the absorption of a neutrino with mass $m_i$:
\begin{eqnarray}
    \frac{d P_{\text{exc},i}}{d \tau} &=& \pi^{-1} |\lambda|^2 |\langle e |\hat{m}_0(0) | g \rangle|^2 \int_{m_i}^\infty d\omega \ \delta(\omega-\Delta E) \nonumber\\\
    &\times& \frac{\omega \sqrt{\omega^2-m^2_i}}{e^{\beta \omega}+1}.
\end{eqnarray}
We see that the above excitation rate is, disregarding the phase-space volume factor, proportional to the mean number of particles with energy $\Delta E$, as it should be. We also note that this detector satisfies the detailed balance condition, relating excitation and absorption rates \cite{CHM08}. Moreover, the excitation rate in this case is also proportional to the expectation value of the (inertial) neutrino density operator, constructed similarly as in Eq.~(\ref{nudensity}), in the state (\ref{inertial_thermal_bath}):
\begin{eqnarray}
    \langle \hat{n}_{i,D} \rangle_{\hat{\rho}_{\beta}} &=& \frac{1}{\pi^2} \int_{m_i}^\infty d\omega_i \frac{\omega_i \sqrt{\omega^2_i-m^2_i}}{e^{\beta \omega_i}+1}.
\end{eqnarray}


\begin{thebibliography}{}
\bibitem{note}
In order to follow the derivation given, e.g.,  in~\cite{CHM08} to write the restriction of $|0_{\text{M}}\rangle$ to the right Rindler wedge in terms of Rindler particle states, we need first to define Rindler modes, similar to $ g^{+ \varpi}_{\vec{k}_{\bot},\sigma}$ and $g^{-\varpi}_{\vec{k}_{\bot},-\sigma}$, on the left Rindler wedge. In this way  we will have a complete ortonormal set of solutions of Dirac's equation in the whole Minkowski spacetime, which can then be related to the inertial modes given in Eq.~(\ref{inertial_modes}) by  the usual Bogolubov transformations.
\bibitem{note3} This statement must be made compatible with situations in which it is sensible to speak of flavor neutrino particles, such as in neutrino oscillation experiments. In most of these experimental situations, the momenta involved are much larger than the neutrino masses $m_i$ and it is useful to define the states given in Eq.~(\ref{flavor_states1}) as coming from weak-processes states \cite{CG} in the limit where there is not enough precision to determine the different neutrino masses. Also, one tacitly assumes enough uncertainty in the momentum determination to allow the neutrinos source to be localized so that neutrino oscillations can be seen \cite{K81}.
\end{thebibliography}
\end{document}